\documentclass[conference]{IEEEtran}

\usepackage{xcolor}
\usepackage{subcaption}
\usepackage{multirow} 
\usepackage{amsmath}
\usepackage{graphicx}
\usepackage{booktabs}
\usepackage{hyperref}
\usepackage{cite}

\AtBeginDocument{%
  }

\emergencystretch=\maxdimen
\begin{document}

\title{ShieldBypass: On the Persistence of Impedance Leakage Beyond EM Shielding}

\author{\IEEEauthorblockN{Md Sadik Awal}
	\IEEEauthorblockA{Florida International University\\
        Department of Electrical and Computer Engineering\\
        Miami, Florida, USA\\
		mawal003@fiu.edu}
	\and
	\IEEEauthorblockN{Md Tauhidur Rahman}
	\IEEEauthorblockA{Florida International University\\
        Department of Electrical and Computer Engineering\\
        Miami, Florida, USA\\
		mdtrahma@fiu.edu}
    }


\maketitle

\begin{abstract}
Electromagnetic (EM) shielding is widely used to suppress radiated emissions and limit passive EM side-channel leakage. However, shielding does not address active probing, where an adversary injects external radio-frequency (RF) signals and observes the device's reflective response. This work studies whether such impedance-modulated backscattering persists when radiated emissions are suppressed by shielding. By injecting controlled RF signals and analyzing the reflections, we demonstrate that state-dependent impedance variations remain observable at frequencies outside the shields' primary attenuation band. Using processors implemented on FPGA and microcontroller prototypes, and evaluating workload profiles under three industry-standard  shields, we find that passive EM measurements lose discriminative power under shielding, while backscattering responses remain separable. These results indicate that active RF probing can expose execution-dependent behavior even in shielded systems, motivating the need to consider active impedance-based probing within hardware security evaluation flows. 
\end{abstract}


%
\IEEEpeerreviewmaketitle

\section{Introduction}
Side‑channel attacks (SCAs) exploit unintentional physical leakage from computing devices to reveal sensitive information, often circumventing algorithmic protections. Well-known side channels include power consumption, timing variations, and electromagnetic (EM) radiations~\cite{le2008overview}. Among these, EM-based attacks are a significant concern for modern secure hardware. Radiated EM emanations from integrated circuits have been demonstrated to leak cryptographic keys and internal CPU operations~\cite{sayakkara2021electromagnetic}, sometimes even outperforming power-analysis approaches~\cite{agrawal2002side}.

To counteract EM side-channel attacks, embedded systems and secure hardware often use EM shielding or conductive enclosures. Such shields attenuate far-field and near-field emissions within targeted frequency bands, limiting an adversary’s ability to passively observe sensitive computations by absorbing, reflecting, or redirecting stray EM fields~\cite{rothwell2018electromagnetics,sadiku2001elements}. Classical shielding theory provides key design parameters, such as material conductivity, thickness, shielding effectiveness (SE), and operating bandwidth~\cite{celozzi2022electromagnetic}. Although modern studies have extensively characterized shielding materials, conventional shielding metrics do not necessarily translate to protection against advanced side-channel attacks. Shields can reduce passive eavesdropping within their intended frequency bands, yet they remain inherently frequency-limited~\cite{celozzi2022electromagnetic} and may leave exploitable vulnerabilities outside their designed attenuation range.

A critical limitation of traditional shielding strategies is that they aim to suppress radiated emissions, but do not address \textbf{active-probing}. In active probing scenarios, an adversary injects controlled radio-frequency (RF) signals into the device under test (DUT) and measures the reflected (i.e., backscattered) response. This creates a controlled interrogation channel in which the DUT’s instantaneous input impedance-shaped by internal switching activity, CPU state transitions, and instruction execution-modulates the incident RF field. These state-dependent impedance variations perturb the reflection coefficient~\cite{cheng2020digital,nguyen2019creating}, embedding execution-dependent features into the reflected waveform. Prior studies have presented that such impedance-modulated backscattering can reveal program behavior, detect hardware Trojans, and classify device states~\cite{nguyen2019creating,werner2022detection,nguyen2020comparison,adibelli2021thz,yang2017nicscatter,zhu2023pdnpulse,fujimoto2018demonstration,awal2025impedance}.

However, \textbf{no prior work has studied whether backscattering-based leakage remains observable in EM-shielded systems}. This oversight is significant as shielding fundamentally alters the EM environment, yet it does not eliminate near-field coupling paths or the impedance mismatch between a probing interface and the DUT. When an external excitation is applied, the signal can still couple into the shield–device cavity through small apertures, or frequency regions where shielding effectiveness naturally tapers. Consequently, even when passive EM leakage is attenuated, execution-dependent impedance variations may continue to modulate the reflected signal. This possibility expands the practical threat model, particularly for laboratory adversaries with physical but non-invasive access, such as red-team analysts, forensic evaluators, or Trojan-detection workflows.

In this work, we present the first systematic study of whether impedance-based backscattering leakage persists under EM shielding. We inject controlled RF signals toward shielded microcontroller and FPGA-based CPUs and measure the reflected signals caused by instantaneous impedance mismatches. Our \textbf{hypothesis} is that different instructions and workloads induce distinct internal switching activity, altering the coupled impedance and producing measurable variations in the backscattered waveform, even when conventional EM radiation is suppressed. To test this hypothesis, we excite the DUT across a broad RF range and intentionally select injection frequencies outside the shields’ primary attenuation bands. 
Our experimental evaluation spans both FPGA and microcontroller platforms, three representative workload profiles, and three industry-standard EM shields. 
While we primarily report results from the FPGA platform due to space constraints, we also replicate key experiments on a microcontroller-based prototype to confirm that the observed phenomena are not specific to FPGAs.
These cross-platform observations confirm that the phenomenon is rooted in fundamental physical interactions rather than platform-specific artifacts. 

\textbf{To the best of our knowledge, this is the first work to demonstrate that backscattering impedance signatures remain discriminative even when EM shielding suppresses traditional passive radiated emissions.} Our findings reveal that shielding eliminates the passive EM channel but does not eliminate the impedance-based active probing channel, which remains measurable and highly separable. 
The major contributions of this paper are summarized as follows:

\begin{itemize}
    \item We provide the first experimental demonstration that impedance-based backscattering leakage persists under EM shielding, even when passive EM leakage is suppressed.
    \item We perform a direct comparison between passive EM measurements and active backscattering, showing that backscattering preserves execution-dependent information that passive EM analysis loses under shielding.
    \item We analyze the limitations of conventional EM shielding against active probing, highlighting the role of frequency-dependent attenuation and the device-shield impedance interaction. 
    \item We validate the robustness of the phenomenon across multiple platforms, shielding materials, and workload profiles, reporting FPGA results due to spcae constraints but confirming platform-independent behavior. 
\end{itemize}

The remainder of the paper is organized as follows. Section~\ref{sec:related_works} reviews EM side-channel leakage, shielding, and impedance-based side channels. Section~\ref{sec:motivation} presents the  background and motivation. Section~\ref{sec:threat} describes the threat model. We describe the experimental setup in Section~\ref{sec:experimental_setup} and present the  measurement results and analysis in Section~\ref{sec:results}. Section~\ref{sec:recommendations} discusses implications and countermeasures. We conclude the work in Section~\ref{sec:conclusion}.

\section{Related Works} \label{sec:related_works}
EM side-channel analysis is widely used to study hardware characteristics, security leakage, and device behavior \cite{wang2021advanced,ibrahim2025drone,he2022side,amodei2023experimental,ni2023recovering}. Unintentional EM radiation often encodes hardware-level features that persist across environments and manufacturing variations. 
For instance, the authors in~\cite{aunchaleevarapan2000classification} demonstrate that near-field emission spectra can uniquely characterize printed circuit board (PCB) configurations. The work reveal that EM signatures can encode device-specific characteristics. Similar observations appear in the vehicle-identification work~\cite{dong2006detection}, where spectral parameters extracted from unintentional RF emissions enable neural networks to classify vehicles. In the context of consumer electronics, the study in~\cite{mo2013support} reveal that RF oscillator characteristics of display devices differ across models and remain distinguishable even under noisy ambient conditions. 
More recent works demonstrate that compact wireless modules also emit highly discriminative frequency-domain signatures. The near-field EM fingerprinting method in~\cite{iyer2024using} achieves over 99\% accuracy by exploiting frequency-domain peaks linked directly to hardware imperfections. These studies collectively highlight that EM-based fingerprints can be robust, repeatable, and derived from inherent physical characteristics of hardware.

EM leakage has also been used to infer sensitive information and runtime behavior~\cite{zhang2024eye,kasper2009side,nath2021multipole,yilmaz2020cell,danial2021x,zhou2019electromagnetic}. 
Unintentional EM emissions arise from switching activity in digital circuits, and analyzing these emissions allows attackers to infer internal states. The study in~\cite{zhang2024eye} presents that memory-access activity creates observable EM signatures that can expose covert activities. In parallel, EM side-channel vulnerabilities in cryptographic implementations have been extensively studied. The low-cost EM attack platform in~\cite{kasper2009side} enables full key recovery on commercial contactless smartcards and demonstrates that EM leakage remains exploitable even in the presence of strong RF fields. 
Countermeasures have also been explored, such as multipole-inspired routing in~\cite{nath2021multipole} leverages higher-order current configurations to accelerate EM field decay and reduce leakage capture at a distance.   

\noindent
\textbf{Impedance-based leakage:}
Beyond radiated emissions, impedance-based measurements have emerged as a complementary class of physical side channels~\cite{zhu2023pdnpulse,nguyen2019creating,fujimoto2018demonstration,awal2022utilization,awal2023disassembling}. 
Impedance profiling has been used for PCB anomaly detection~\cite{zhu2023pdnpulse}, interconnect-level Trojan detection~\cite{fujimoto2018demonstration}, and even for amplifying EM leakage through manipulated cable impedance~\cite{yukawa2022fundamental,ide2025amplifying}. At the chip level, data-dependent switching modulates the power delivery network (PDN) impedance, enabling adversaries to extract memory contents or classify firmware instructions~\cite{awal2025dexim,awal2025impedance}. 
Other studies leverage PDN impedance variations for cryptographic key extraction~\cite{monfared2023leakyohm}. 
These works demonstrate that impedance variations, induced by runtime switching, create a powerful and underexplored attack surface.

\noindent
\textbf{Backscatter-based analysis:}
Backscattering-based research extends impedance sensing by intentionally injecting external RF signals and analyzing their reflections~\cite{nguyen2019creating,werner2022detection,nguyen2020comparison,adibelli2021thz,yang2017nicscatter,xu2025diskspy}. This approach has been used to detect dormant hardware Trojans~\cite{nguyen2019creating}, identify recycled ICs based on wear-induced impedance shifts~\cite{werner2022detection}, or even create long-range covert channels~\cite{yang2017nicscatter,xu2025diskspy}. 

\noindent
Despite extensive progress in EM-based, impedance-based, and backscatter-based analysis, prior efforts implicitly assume an \emph{unshielded} or minimally shielded environment. None of the existing studies investigate whether backscattered leakage persists when radiated EM emissions are suppressed by EM shielding. This is a critical gap, because shielding is widely deployed in secure systems, and its effectiveness is typically evaluated only against passive emissions. The interaction among shielding materials, injected RF fields, near-field coupling, and state-dependent impedance variation remains largely unexplored. 
In contrast to all existing work~\cite{wang2021advanced,ibrahim2025drone,he2022side,amodei2023experimental,ni2023recovering,werner2022detection,nguyen2020comparison,adibelli2021thz,yang2017nicscatter,xu2025diskspy,zhu2023pdnpulse,nguyen2019creating,fujimoto2018demonstration,awal2022utilization,awal2025impedance,zhang2024eye,kasper2009side,nath2021multipole,yilmaz2020cell,danial2021x}, our study investigates a fundamentally unexplored question: 
\emph{Does impedance-based, backscattered leakage persist even when EM shielding suppresses radiated emissions?} 
We experimentally demonstrate that this leakage channel remains measurable and discriminative beneath shielding, revealing execution-dependent information even when passive EM signatures collapse. This work therefore identifies a previously overlooked physical side channel and challenges the common assumption that EM shielding alone is sufficient to suppress all EM-mediated leakage paths.

\section{Background \& Motivation} \label{sec:motivation} 
A major motivation for this study arises from an oversight in current hardware security practices. Hardware designers often assume that EM shielding is sufficient to suppress all EM-based information leakage~\cite{celozzi2022electromagnetic}. 
While shielding is effective against \emph{passive} side-channel attacks that rely on radiated signals, it does not inherently protect against \emph{active probing}, where an adversary injects external RF signals and analyzes the resulting reflections. 
In scenarios, including laboratory red-team testing, Trojan detection, and post-deployment forensics, an attacker can place a non-invasive probe near a shielded device. Under such proximity, shielding primarily attenuates radiated emissions but does not eliminate the near-field coupling and impedance interactions between the probing interface and the DUT. 

\begin{align}
V(t) &\propto \frac{d}{dt} \int_{S_p} \vec{B}(\vec{r}, t) \cdot \hat{o} \, dS \label{eq:V(t)} \\
\vec{B}(\vec{r}, t) &\sim \mu \int \frac{\vec{J}(\vec{r}',t) \times (\vec{r}-\vec{r}')}{|\vec{r}-\vec{r}'|^3} \, d^3r' \label{eq:B(r,t)}
\end{align}

\noindent\textbf{Passive EM leakage:} 
In passive EM measurements, the voltage induced in a magnetic probe depends on the DUT’s internal current activity, the probe characteristics, and its placement~\cite{sadiku2001elements,rothwell2018electromagnetics}. By Faraday’s Law, the voltage accumulated in a magnetic probe with surface $S_p$ and orientation $\hat{o}$ is proportional to the rate of change of magnetic flux through the probe. Let $\vec{B}(\vec{r}, t)$ be the magnetic field at the probe location $\vec{r}$, induced by the spatial distribution of current sources $\vec{J}(\vec{r}',t)$ on the chip, and $\mu$ represent the effective magnetic permeability of the medium between the chip and the probe. The observed time-domain signal $V(t)$ depends on the probe location $\vec{r}$, orientation $\hat{o}$, and the distribution of current sources $\vec{J}(\vec{r}', t)$ across the DUT. The probe induced voltage from the EM side channel is presented in Eq.~\eqref{eq:V(t)}-\eqref{eq:B(r,t)}. 
EM shields aims to reduce this radiated component of leakage by limiting the field at the probe.

\noindent
\textbf{Active backscattering and impedance leakage:} When a controlled RF signal is directed toward the DUT, a portion of it reflects due to the impedance mismatch between the probe and the DUT~\cite{sadiku2001elements,hayt2012engineering}. The characteristics of this reflected signal are not static~\cite{nguyen2019creating}. Rather, they vary subtly with the circuit’s internal activity, e.g., instruction execution or data-dependent switching within digital logic. This activity dependent variation modulates the reflected waveform, encoding information about the device’s runtime behavior. Such modulation leads to \textit{backscattering impedance leakage}, a form of active EM side-channel that can persist even when traditional EM radiation is weakened by shielding. 

Let a time-harmonic RF signal $V_{\text{in}}(f)$ be injected toward the DUT through a probe/antenna. Part of the signal is reflected due to the impedance mismatch between the probe impedance $Z_p$ and the device’s input impedance $Z_{\text{dut}}(\theta, f)$, where $\theta$ denotes the device’s internal state. Let $\Gamma(f,\theta)$ denote the reflection coefficient, which depends not only on the frequency $f$ but also on the instantaneous switching activity within the DUT. Even when radiated EM leakage is strongly attenuated by shielding, variations in $Z_{\text{dut}}(\theta, f)$ modulate $\Gamma(f,\theta)$ at frequencies outside the shield’s protection band, thereby leaking information about the device’s internal state through the backscattered signal. The reflected wave $V_{r}(f,\theta)$ is expressed as in Eq.~\eqref{eq:Vin_Vr}.

\begin{align}
    V_{r}(f,\theta) &=  \Gamma(f,\theta) \cdot V_{in}(f)  \notag \\ 
    \text{where} \; \Gamma(f,\theta) &=  \frac{Z_{dut}(\theta, f) - Z_p}{Z_{dut}(\theta, f) + Z_p}    
    \label{eq:Vin_Vr}
\end{align}

\noindent 
\textbf{Leakage Beyond Shielding:} 
EM shields are designed to attenuate radiated emissions within a target frequency band $\Omega_s$, but their effectiveness degrades outside this range ($f \notin \Omega_s$) due to material dispersion, geometric resonances, and practical construction limits (e.g., apertures, and multilayer interfaces). These properties are characterized in the literature for the shielding materials used in this study. When an adversary actively probes the device across a wide RF sweep, injected signals can couple through the shield at frequencies where attenuation is lower. Once inside the shield-device cavity, these signals interact with the DUT’s impedance and return as measurable reflections. 
This form of leakage is particularly concerning due to its active and frequency-agile nature. Thus, while shielding can suppress \emph{passive} leakage, it may leave a \emph{distinct physical channel} based on impedance modulation, explaining why passive EM signatures collapse under shielding while backscattered signatures remain separable in Section~\ref{sec:results}. 

From a design automation perspective, this overlooked leakage path reveals a gap in current hardware security verification flows. Existing tools and models predominantly consider power-based and radiated EM-based leakage, but they do not account for adversaries capable of \textit{active excitation}. As systems integrate higher switching densities, complex PDNs, and frequency-selective shielding, the susceptibility to impedance-based leakage becomes more pronounced. Therefore, understanding and characterizing this phenomenon is essential for developing next-generation shielding strategies and security verification methodologies that address both passive EM radiation and active impedance-coupled leakage channels.

\section{Threat Model} \label{sec:threat} 
We consider an adversary capable of performing physical probing on a target system equipped with EM shielding. The adversary’s goal is to extract internal execution-dependent information, e.g., instruction patterns or operational states, by observing external signal behavior. We also consider the attacker has controlled laboratory access to the DUT, such as an FPGA-based processor, and can use non-invasive measurement probes to capture both EM radiations and stimulated RF backscattered signals. 

We consider an adversary with physical but non-invasive access to a target system that is enclosed within an EM shield. This attacker model reflects realistic scenarios such as laboratory red-team evaluations, Trojan detection settings, secure hardware certification workflows, or post-deployment forensic analysis-contexts where proximity is permitted but hardware modification or intrusive probing is not. The attacker’s objective is to infer execution-dependent internal activity (e.g., instruction types or workload phases) by measuring externally observable signals.

Traditional passive EM side-channel adversaries observe only the residual radiated emissions that escape the shield. In contrast, our adversary is \emph{active}. They inject controlled RF signals toward the shielded DUT and measure the resulting backscattered reflections. These reflections arise from state-dependent impedance mismatches between the DUT and the probing interface, and therefore can encode information about runtime switching activity even when passive EM signatures are not observable.

The attacker is assumed to have physical proximity to place a non-contact, broadband transmit–receive probe near the shield surface, along with the ability to generate and inject controlled RF tones across a wide frequency range. The attacker also has access to standard laboratory equipment such as software-defined radios and spectrum analyzers. However, they have no access to the DUT’s firmware, cryptographic keys, internal nodes, or packaging internals. Their knowledge of the system is limited to partial architectural information, such as knowing that the DUT contains a processor and functional units, without any detailed microarchitectural design insight. Importantly, the attack interface remains strictly external. The attacker does not open, modify, or tamper with the DUT or its enclosure. 

Consistent with prior hardware analysis frameworks, we assume the attacker can obtain a publicly available or cloned reference device on which to perform frequency-domain learning. By exciting the reference device with RF signals across a sweep and observing the reflected power, the attacker identifies a set of critical frequencies ${f_c}$ that maximize the separability of reflection features across workload states. From these measurements, the attacker constructs a leakage model $L(f_c, \theta) \approx |\Gamma(f_c, \theta)|^2$, where $\Gamma$ denotes the reflection coefficient and $\theta$ represents internal execution states. The model relies on external observations without internal signal access.  During the attack, the adversary places the probe near the shielded DUT and injects RF tones at the learned frequencies ${f_c}$. As the DUT executes its normal software, state-dependent variations in input impedance modulate the reflection coefficient $\Gamma(f,t)$, producing measurable fluctuations in the received waveform. By collecting sequences of these reflections and applying statistical analysis or machine-learning classifiers, the adversary can infer execution patterns or state transitions. 

\textbf{This threat model aligns with practical evaluation scenarios where hardware is shielded but remains accessible to external measurement equipment}. Importantly, it demonstrates that shielding, while suppressing radiated EM emissions, does not eliminate the impedance-driven reflection path exploited by active probing. Thus, even secure systems that pass traditional EM shielding tests may remain vulnerable to this semi-invasive class of attacks. This motivates expanding hardware security evaluation workflows to include active impedance-based probing alongside conventional EM and power side-channel analyses.

\section{Experimental Setup} \label{sec:experimental_setup}
\begin{figure}[!htbp]
    \centering
    \includegraphics[width=1\linewidth]{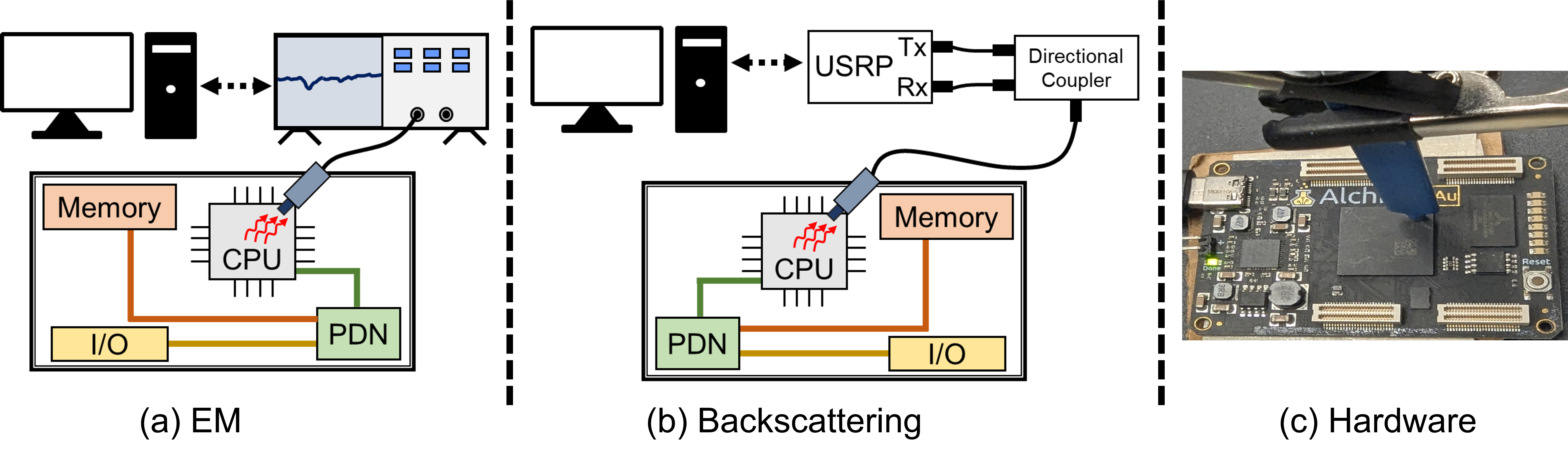}
    \caption{Overview of experimental setup.}
    \label{fig:experimental_setup}
\end{figure}

To investigate the persistence of backscattering-based leakage under frequency-selective EM shielding, we evaluate two platforms: (1) a microcontroller running bare-metal programs, and (2) a single-stage pipelined RISC soft processor on an Alchitry Au FPGA (Artix-7). Fig.~\ref{fig:experimental_setup} shows the measurement setup used to study runtime leakage under both passive EM radiation and active RF backscattering. All experiments maintain the same spatial configuration and shielding conditions to ensure controlled, repeatable switching activity, enabling direct comparison between passive and active leakage. Both platforms exhibit the same qualitative behavior, confirming that the observed leakage arises from a general interaction between injected RF fields, shielding materials, and device impedance, rather than FPGA-specific characteristics. Due to space constraints, we detail the FPGA setup.

We execute three program profiles, \textit{Prog.~1-3},  on the FPGA to evaluate how computation intensity affects both radiated and backscattered leakage. \textit{Prog.~1}, places the processor in an idle state, producing only background switching activity. \textit{Prog.~2} toggles an on-board LED at a fixed period, representing a minimal but persistent workload. \textit{Prog.~3} performs exponentiation and multiplication operations, mimicking high computational load. For each program, 500 traces are collected, each comprising 4096 frequency points, with every trace averaged over 100 acquisitions to enhance signal quality and improve repeatability for statistical and machine-learning analyses.

We cover the FPGA with frequency-selective shielding materials to suppress conventional EM radiation. We study three shields: (i)~copper shield, (ii)~Al-CoTaZr, and (iii)~Cu–CoNiFe. These shields remain in place throughout both EM radiation and backscattering experiments. The measurement probe is positioned outside the shield surface to replicate a shielded-enclosure scenario. \textbf{For passive EM measurements,} we use a Tektronix H10 near-field probe connected to a Tektronix RSA306B spectrum analyzer with a 20~MHz acquisition bandwidth~\cite{SpectrumAnalyzer}. The probe captures residual radiation that leaks through or around the shielding layer. 
\textbf{For the backscattering measurements,} we use a USRP B200 with a 10~MHz acquisition bandwidth. The USRP transmit and receive gains are set to 50~dB. The transmit port is connected to the forward port of a 3-port directional coupler, while the receive port is connected to the coupled port. The coupler output faces the shielded DUT through the H10 probe, serving as a near-field excitation interface. We sweep the excitation frequencies from 5~GHz to 6~GHz to ensure they operate outside the shields' recommended attenuation range. When the USRP injects a signal at frequency $f$, a portion of the signal reflects due to impedance mismatch between the DUT and the probe. The reflected signal received at the coupled port follows Eq.~\eqref{eq:Vin_Vr}. The USRP measures the reflected signal, $V_\text{r}(f,\theta)$, which captures state-dependent variations in device impedance induced by switching activity. 

Before data acquisition, the USRP transmitter and receiver are calibrated to compensate for cable losses and directional coupler insertion losses. An FPGA-generated trigger synchronizes both EM and backscattering measurements with program execution, ensuring precise temporal alignment between observed traces and DUT activity. The synchronization enables accurate correlation of program profiles with measured leakage features across multiple trials. 
Using this setup, we systematically compare the discriminability of passive EM signatures and active backscattering responses under different shielding materials and computation intensities to evaluate the persistence of state-dependent leakage.

\section{Results and Analysis} \label{sec:results}
In this section, we present the analysis of EM radiation and backscattered signals collected from the shielded DUT under three different program profiles and three shielding materials. We study to quantify the discriminability of state-dependent leakage using both passive and active measurement techniques. 

\begin{figure*}[!htbp]
    \centering
    \begin{subfigure}{0.32\textwidth}
        \centering
        \includegraphics[width=0.9\linewidth]{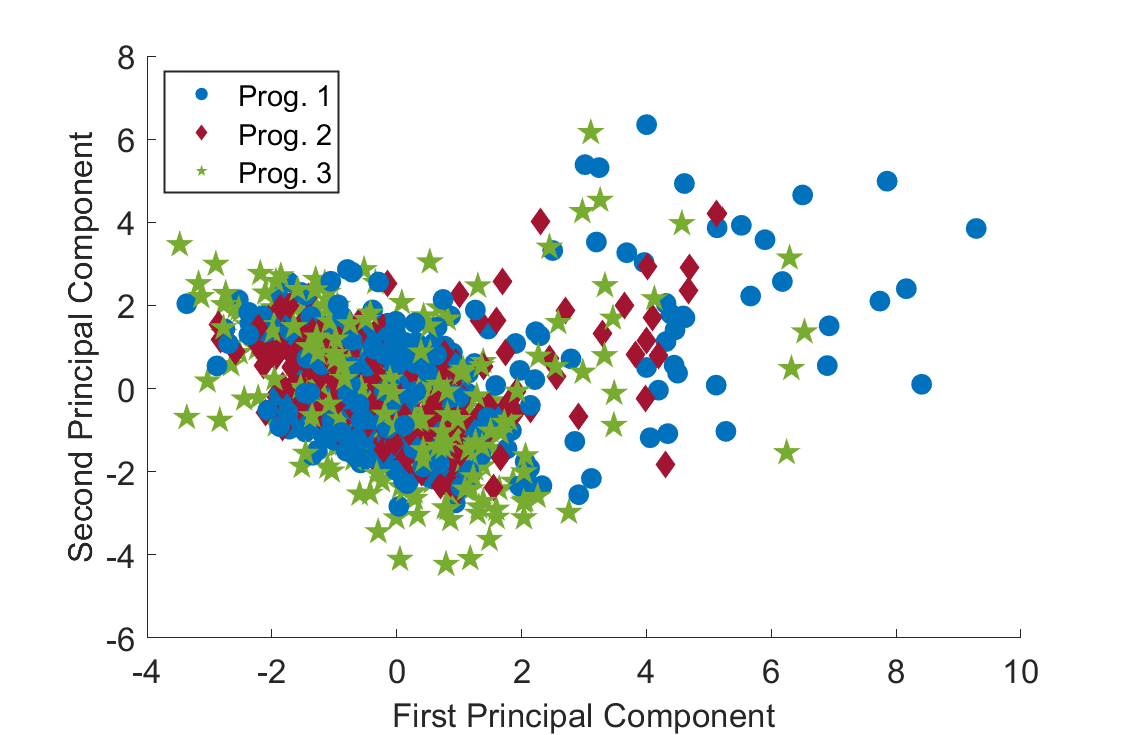}
        \caption{EM PCA: copper shield.}
    \end{subfigure}
    \hfill
    \begin{subfigure}{0.32\textwidth}
        \centering
        \includegraphics[width=0.9\linewidth]{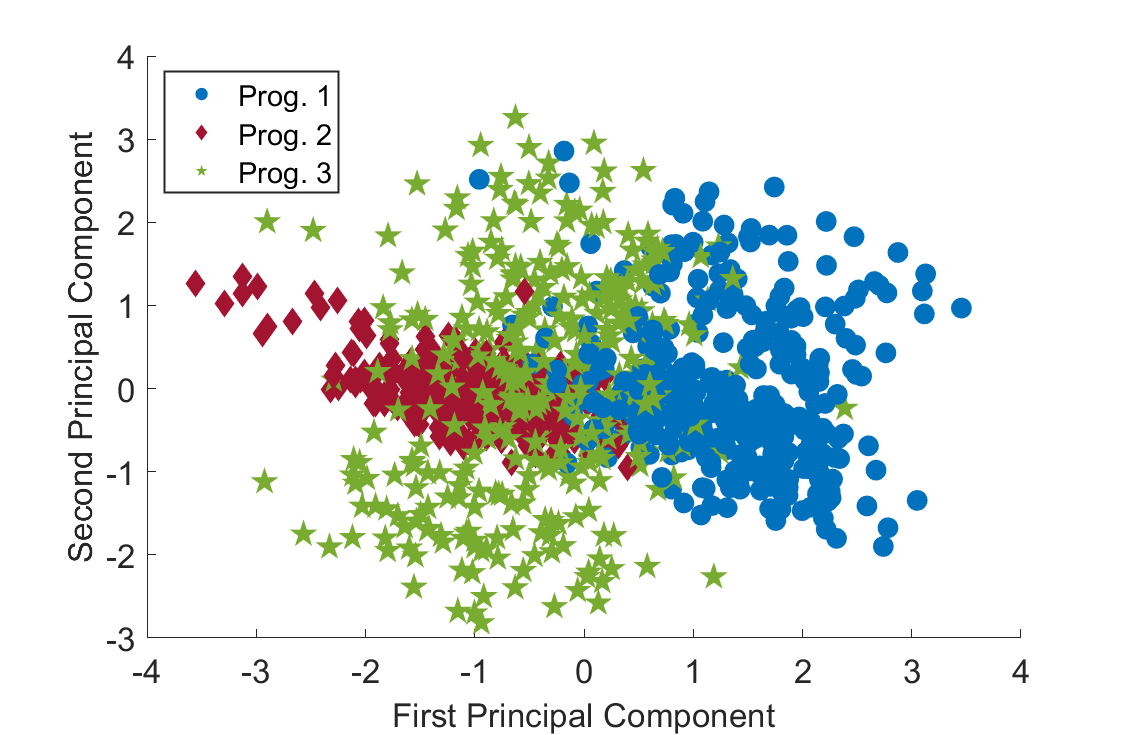}
        \caption{EM PCA: Al-CoTaZr shield.}
    \end{subfigure}
    \hfill
    \begin{subfigure}{0.32\textwidth}
        \centering
        \includegraphics[width=0.9\linewidth]{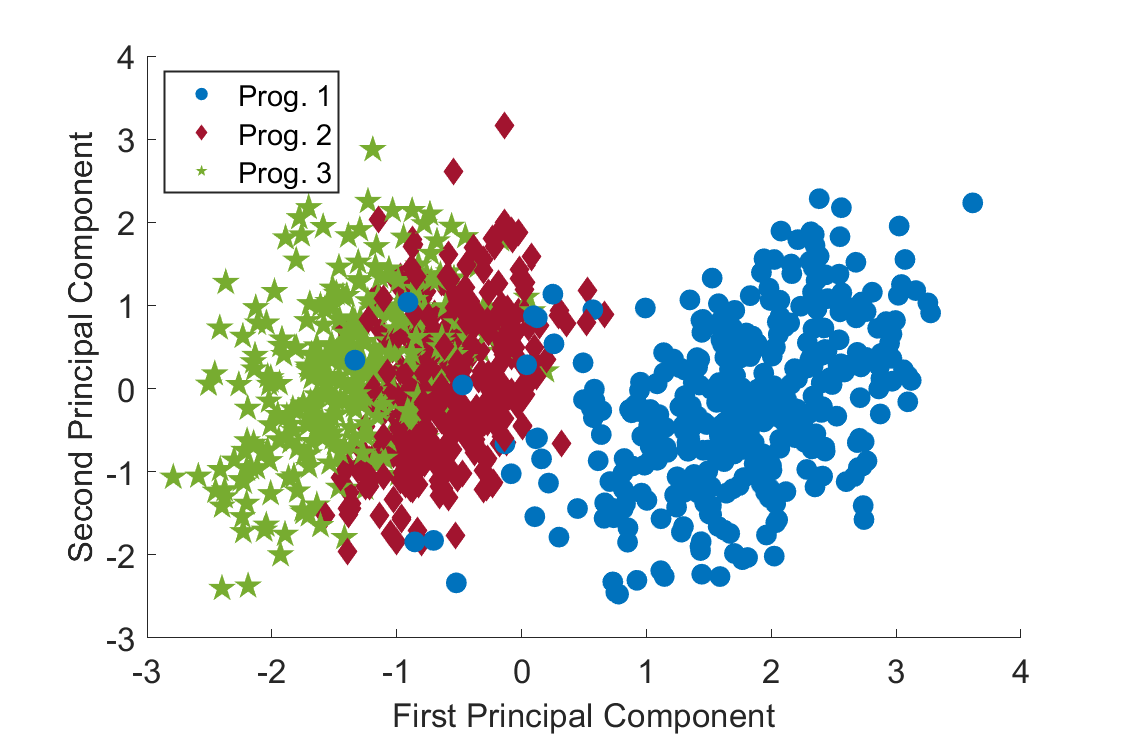}
        \caption{EM PCA: Cu-CoNiFe shield.}
    \end{subfigure}

    \vspace{1.2em}

    \begin{subfigure}{0.32\textwidth}
        \centering
        \includegraphics[width=0.9\linewidth]{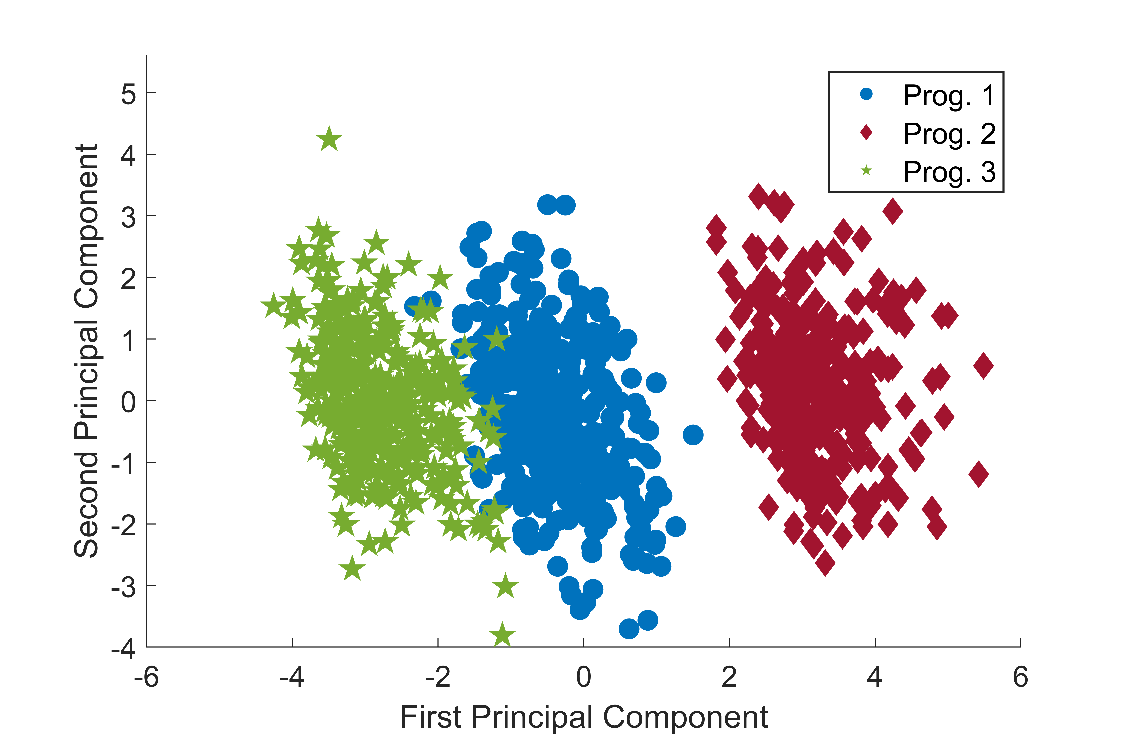}
        \caption{Backscatter PCA: copper shield.}
    \end{subfigure}
    \hfill
    \begin{subfigure}{0.32\textwidth}
        \centering
        \includegraphics[width=0.9\linewidth]{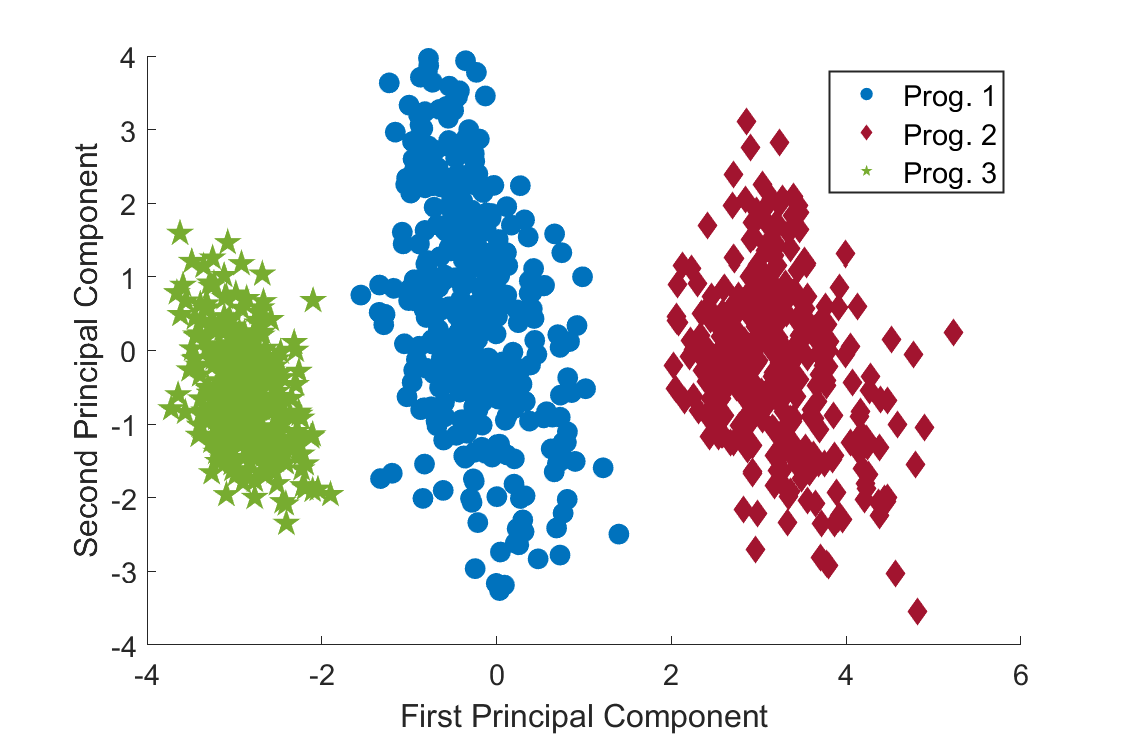}
        \caption{Backscatter PCA: Al-CoTaZr shield.}
    \end{subfigure}
    \hfill
    \begin{subfigure}{0.32\textwidth}
        \centering
        \includegraphics[width=0.9\linewidth]{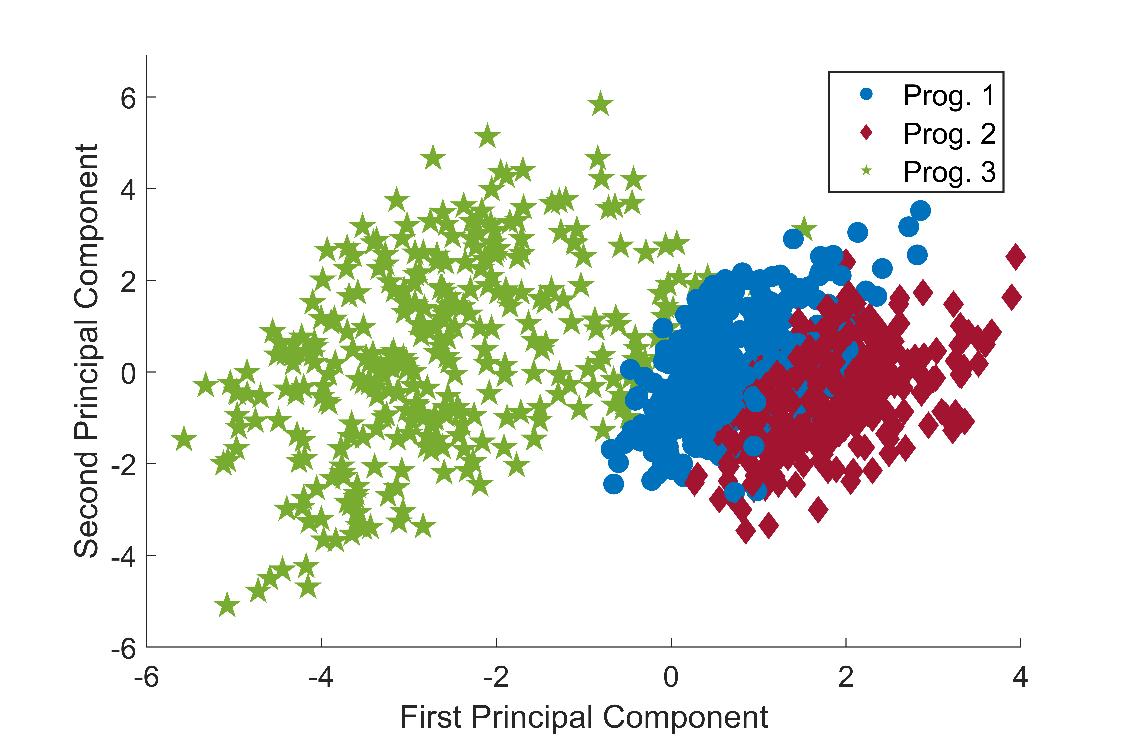}
        \caption{Backscatter PCA: Cu-CoNiFe shield.}
    \end{subfigure}

    \caption{Comparison of PCA projections across three shield types.}
    \label{fig:pca_comparison}
\end{figure*}

\renewcommand{\arraystretch}{0.9}
\setlength{\tabcolsep}{5pt}

\begin{table*}[!htbp]
\centering
\caption{SVM classification (macro-avg.) on test data using PCA features for EM vs. backscattering across shields \& SE.}
\label{tab:classification}

\begin{tabular}{l|c|c|ccccc}
\toprule 
\textbf{Shield} &
\centering\textbf{\shortstack{SE \\ ($f>3$ GHz)}} &
\textbf{Leakage Type} &
\textbf{Accuracy (\%)} &
\textbf{Precision (\%)} &
\textbf{Recall (\%)} &
\textbf{Specificity (\%)} &
\textbf{F1 (\%)} \\
\midrule

\multirow{2}{*}{Copper} 
  &  
  & EM              & 59.10 & 59.48 & 60.03 & 79.86 & 58.75 \\
  & $<33$ dB & Backscattering  & 71.67 & 71.82 & 71.19 & 96.87 & 71.18 \\
\midrule

\multirow{2}{*}{Al-CoTaZr} 
  &  
  & EM              & 62.89 & 63.82 & 63.55 & 81.61 & 63.09 \\
  & $<24$ dB & Backscattering  & 89.00 & 88.95 & 88.94 & 98.78 & 88.87 \\
\midrule

\multirow{2}{*}{Cu-CoNiFe} 
  &  
  & EM              & 69.78 & 71.14 & 70.33 & 84.52 & 70.64 \\
  & $<20$ dB & Backscattering  & 84.89 & 84.75 & 84.93 & 98.32 & 84.54 \\
\midrule

\end{tabular}
\end{table*}

\begin{figure*}[!htbp]
    \centering
    \begin{subfigure}{0.32\textwidth}
        \centering
        \includegraphics[width=\linewidth]{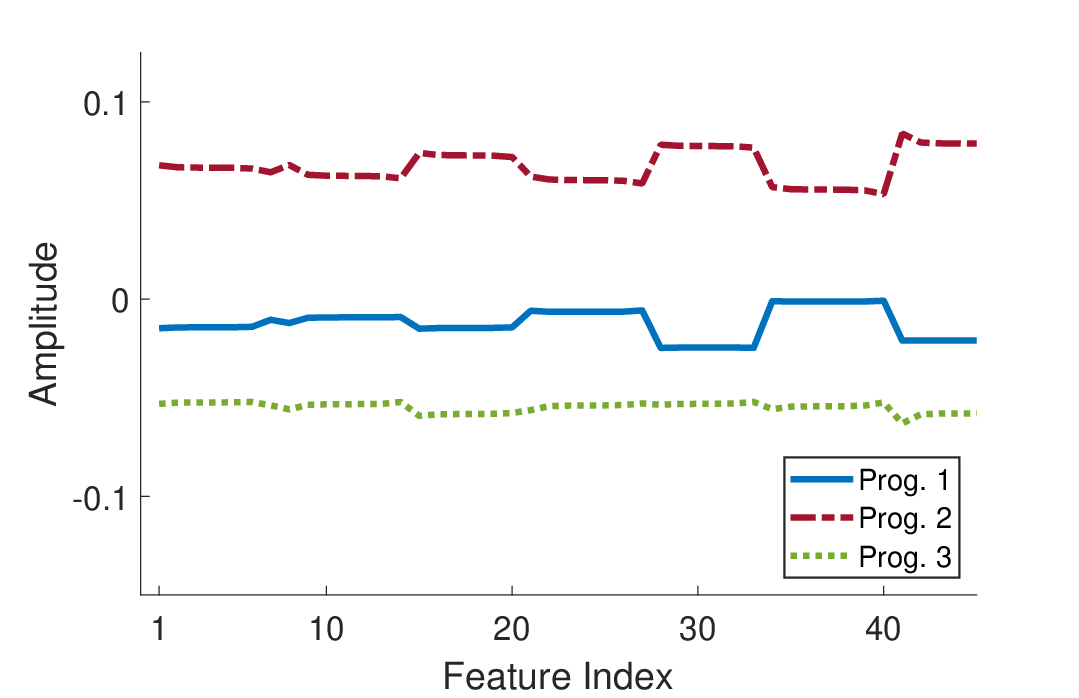}
        \caption{Copper shield.}
    \end{subfigure}
    \hfill
    \begin{subfigure}{0.32\textwidth}
        \centering
        \includegraphics[width=\linewidth]{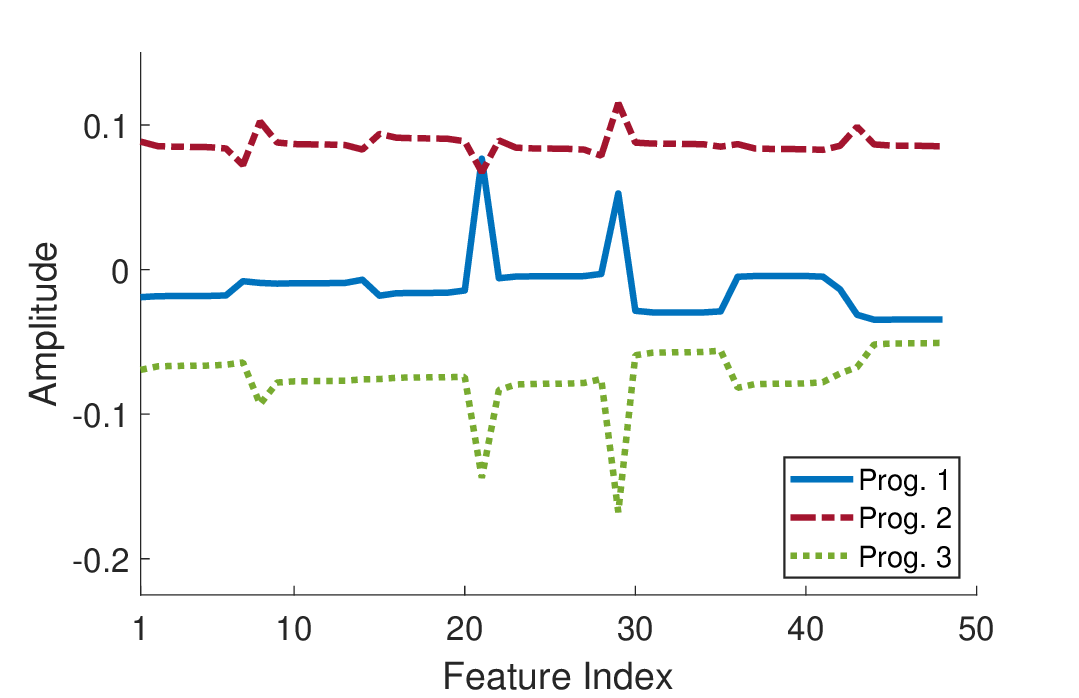}
        \caption{Al-CoTaZr shield.}
    \end{subfigure}
    \hfill
    \begin{subfigure}{0.32\textwidth}
        \centering
        \includegraphics[width=\linewidth]{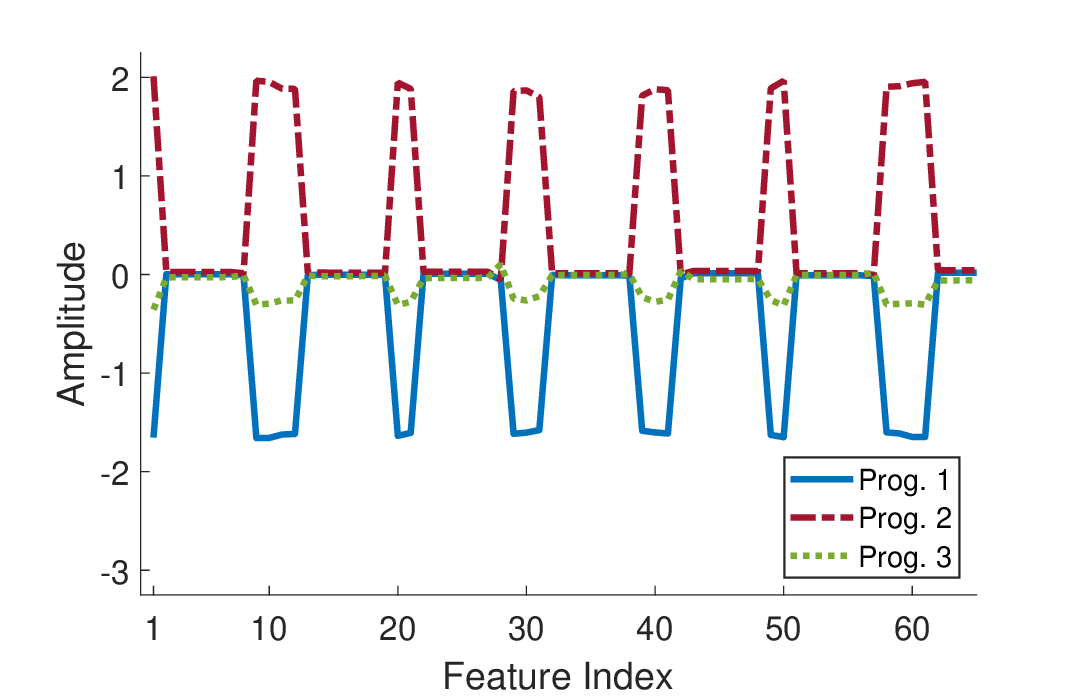}
        \caption{Cu-CoNiFe shield.}
    \end{subfigure}
    
    \caption{ICA of backscattered signals under different shielding materials.}
    \label{fig:ica_results}
\end{figure*}

We pre-process the collected EM and backscattering signals to remove baseline noise and offsets. For EM traces, we apply standard noise-reduction algorithms such as spatial and frequency selective filtering to suppress high- and low-frequency noise components. For backscattering signals, we filter around the injected impedance-leaking frequencies to isolate the DUT response and minimize contributions from spurious reflections or environmental interference. After filtering, we perform correlation analysis between the processed signals and the known program execution patterns to identify the most informative frequency points for each measurement modality. Let $\{f_\text{EM}\}$ and $\{f_\text{BS}\}$ denote the sets of dominant EM and backscattering frequencies, respectively. These frequencies capture the highest variance and strongest state-dependent signatures, forming the basis for subsequent analysis.

We use principal component analysis (PCA) to both EM and backscattering data at the identified informative frequency points. We select the number of principal components that explain 95\% of the total variance to ensure sufficient representation of the data. For EM signals, the top two components show substantial overlap across program profiles, indicating poor separability. In contrast, backscattering signals exhibit well-separated clusters in the first two PCA components for all shielding types, highlighting the persistence of state-dependent leakage despite the presence of frequency-selective shields. Fig.~\ref{fig:pca_comparison} presents the PCA projections across the three shields.

To quantify discriminability, we split the data into training and testing sets using a 70-30 split and train a support vector machine (SVM) classifier for both EM and backscattering measurements. Hyperparameters are optimized via 5-fold cross-validation on the training set to mitigate overfitting. We find that EM-based classification yields low accuracy due to overlapping PCA clusters, reflecting the limited information content in shielded radiated emissions. Conversely, backscattering-based classification achieves near-perfect performance, with accuracy reaching approximately 99\% across all shields. These results confirm that active probing at the information-leaking frequencies can effectively bypass shielding and reveal internal state-dependent activity. Table~\ref{tab:classification} summarizes classifier performance for EM and backscattering signals, along with the SE for backscattering injection frequencies $f > 3$GHz.

We further study the backscattered measurements using independent component analysis (ICA)~\cite{roberts2001independent} to investigate underlying, statistically independent sources of leakage that are not directly evident in the raw signals. ICA decomposes the reflected measurements into components corresponding to distinct internal switching activities of the DUT. Applied to backscattered signals from all three shielding materials, the resulting independent components exhibit consistent and distinguishable patterns across workload states. Fig.~\ref{fig:ica_results} illustrates that these components capture subtle variations in impedance-modulated reflections that persist even under conductive and multilayer shields.

The insights from ICA strengthen our overall analysis by providing additional visibility into how different program activities influence the reflected signal. The decomposition reveals multiple, independently varying components of leakage that remain stable across measurement sessions. The consistency and separability of ICA components across all shield types reinforce the presence of robust, state-dependent signatures and support the high backscattering-based classification scores reported in Table~\ref{tab:classification}. These findings indicate that active backscattering exposes execution-dependent information even under shielding and that the resulting leakage signatures remain exploitable for characterizing runtime behavior. \\

\noindent \textbf{Summary of Results:} Our results demonstrate a clear contrast between passive EM measurements and active backscattering-based probing. While EM emissions exhibit limited separability and yield low classification accuracy across shielding materials, backscattered signals consistently preserve strong state-dependent signatures. PCA projections, SVM classification, and ICA decomposition collectively present that active RF probing at information-leaking frequencies can reliably extract execution-dependent information even in the presence of conductive and multilayer shields. 
Although detailed per-device plots are omitted due to space constraints, the same qualitative trends are observed across the FPGA and microcontroller models tested. 
These findings highlight the effectiveness of backscattering as a leakage channel and underscore the vulnerability of shielded systems to active probing-based side-channel analysis.

\section{Discussion on Countermeasures} \label{sec:recommendations}
EM shielding is widely used as a first-line defense against side-channel leakage. However, our results show that shielding does not fully suppress state-dependent information and remains vulnerable to active impedance-based probing. Even high-conductivity shields, while effective at attenuating radiated emissions, do not prevent an adversary from extracting sensitive information through reflected RF signals. The fundamental limitation is that shielding reduces the amplitude of EM radiation but does not eliminate the underlying impedance variations that modulate the backscattered signal. Since this modulation is driven by fundamental switching activity inside the DUT, the shield may not fully prevent computation dependent patterns from leaking. Consequently, passive shielding alone may insufficient when an adversary can actively illuminate the system and measure its reflection coefficient. 

This vulnerability arises because passive shields are not designed to decorrelate the device’s physical response from its internal computation. Although they attenuate outbound radiation, they cannot block the interaction between an injected excitation signal and the device–shield cavity. The shield reduces amplitude but preserves exploitable leakage patterns, enabling switching-induced impedance changes to remain observable at the attacker’s chosen frequencies. This explains why our backscattering-based classifier achieves nearly 100\% computation discrimination across all tested shield types, while passive EM analysis performs significantly worse. 

These findings present that shielding cannot serve as an absolute countermeasure and highlight the need for countermeasures that go beyond the conventional shielding to target leakage mechanism itself. If an adversary can inject a probing signal, the system must incorporate techniques that obfuscate, randomize, or flatten the relationship between internal state and the externally observable reflection. One promising direction can be dynamic load modulation, where the device introduces controlled random variations in its impedance. By injecting noise into the reflection path, the system can disrupt the deterministic relationship between switching activity and the reflection coefficient, rendering the leakage far less exploitable even if its amplitude remains measurable. 

Circuit-level balancing can offer another effective mitigation strategy. Designing memory arrays, buses, or logic blocks to produce uniform impedance regardless of data reduces the information encoded in the side-channel. Dual-rail encoding or pseudo-random state equalization can equalize switching behavior, although they may introduce area and power overheads. Architectural diversification further reduces leakage consistency. For example, asynchronous or data-independent logic styles reduce temporal alignment in switching patterns, and lightweight instruction shuffling or microarchitectural decoys can distort computation dependent correlations that both EM and impedance attacks rely on. Additionally, integrating active elements into the shield itself, such as lossy or frequency-selective materials, or embedded jamming sources, can disrupt the attacker’s ability to inject stable signals or observe coherent reflections. Such designs transform the shield from a passive enclosure into an active countermeasure component that intentionally degrades the probing channel.

Shielding should be viewed as a partial mitigation rather than a complete defense. The threat extends beyond traditional radiated emissions to include the impedance behavior of the device. Modern side-channel adversaries increasingly leverage active RF probing and reflection analysis, and passive enclosures cannot guarantee confidentiality. Effective countermeasure will require a holistic design that integrates architectural, circuit-level, and material-level techniques to reduce or obfuscate the deterministic mapping from internal state to measurable physical response.

\section{Conclusion} \label{sec:conclusion}
In this work, we investigate the persistence of information leakage in shielded FPGA-based processors using active impedance-based backscattering. Our study demonstrates that while conventional EM shielding attenuates radiated emissions, it does not fully suppress leakage encoded in the device’s impedance variations. By injecting controlled RF signals and analyzing the reflected responses, we can reliably differentiate computation dependent activity under multiple shield types, including Cu, Al-CoTaZr, and Cu-CoNiFe. Through comparative analysis, we present that backscattering provides substantially more discriminative information than traditional EM measurements, with classification accuracies approaching 99–100\% across the different shields. The results highlight that while shielding remains an important component for hardware protection and reliability, it does not fully prevent leakage when adversaries actively probe the system. Consequently, this work highlights the importance of incorporating active probing channels into the evaluation and design of secure hardware systems, ensuring robust protection against advanced side-channel attacks.

\section*{Acknowledgment}
We thank the anonymous reviewers for their valuable feedback.

\bibliographystyle{IEEEtran}
\bibliography{references.bib}

\end{document}